# An Analysis of the Impact of SEO on University Website Ranking[A]

Mohammad Javad Shayegan

Department of Computer Engineering, University of Science and Culture, Tehran, Iran

shayegan@usc.ac.ir

Maasoumeh Kouhzadi

Department of Computer Engineering; Maybod Branch; Islamic Azad University
Maasoumeh.kouhzadi@gmail.com

**Abstract**

Today, ranking systems in universities have been considered by the academic community and there is a tight competition between world universities to achieve higher ranks. In the meantime, the ranking of university websites is also in the spotlight and Webometric research center announces the ranks of university websites twice a year. Examining university rankings indicators and Webometric ranks of the university indicates that some of these indicators directly and indirectly affect each other. On the other hand, a preliminary study of Webometric indicators shows that some Search Engine Optimization (SEO) indicators can affect Webometric ranks. The purpose of this research is to show how far the SEO metrics can affect the website rank of the university. To do this, after extracting 38 points of the significant SEO metrics of the extracted universities using various tools, data analysis was conducted along with applying association rules on the data. The results of the research show that some of the SEO metrics, such as the number of back links, Alexa Rank and Page Rank have a direct and significant impact on the website rank of universities and in this regard, interesting rules have been extracted.

## 1. Introduction

Currently, the evaluation and ranking of university websites is as important as the evaluation and ranking of educational and research activities of universities. Ranking more than 24,000 universities in the world has been a project since 2004 by the cybermetric laboratory of Spain[1] which known Webometric[B]. There are currently four main parameters as Webometric evaluation indicators. Adherence to SEO metrics or optimization for search engines in the technical and content implementation of academic websites is a very important issue that will affect all four ranking parameters. From millions of websites on the Internet, users usually focus on the first pages of search engine results[2]. According to the studies, websites of reputable universities consider many parameters that are important for search engines and optimize their websites accordingly. Therefore, this study aims to investigate the impact of SEO parameters on university ranks. Moreover, evaluation and measurement of the university websites in this regard can present the strengths and weaknesses of the websites to their administrators and designers and help them design, correct defects, and provide some information to improve the services, attract more audiences, and increase their satisfaction.

---

[B] webometric has a general definition i.e. ranking web of universities.. Webometric is a popular research center for weboetric. The ranking system of the current research, follow the Webometric ranking system.

## 2. Background of the Research

International ranking of universities is an inseparable part of the higher education prospect. However, they focus only on a few hundred universities out of the more than 20,000 higher education institutions worldwide. Based on this classification, the rankings of the universities are different in terms of presence, effectiveness, the level of openness of the Internet sites and their accessibility, and the degree of excellence of the rank[3].

Shi et al conducting a SEO strategy from aspects of directory structure, keyword, strategy, URL pseudo-static, code optimization, and input links are under the study[4]. In a study by Kargar, priorities of 18 extracted metrics were obtained and for a quantitative evaluation, 15 sub-metrics, which can be obtained automatically, were extracted. The results of the study showed that there is a significant relationship between presentation metrics and SEO parameters[5].

In the study of Reddy et al., parameters of website designing standards, search engine optimization, and site indexing in valid search engines in Chinese and Indian universities were studied and compared. From the above-mentioned parameters, it was concluded that Chinese universities are ranked better because they consider these parameters more[6].

Dwenger et al. , in an exploratory analysis of websites links of the universities in Nigeria which was performed using the AltaVista search engine, observed that university websites in the country have good link relations with each other and most of their internal and external links are at the national level, but a small percentage of the links are related to university websites outside Nigeria[7]. Moreover,in the study of Ortega et al. the relationship between universities in Europe and between European countries was studied for their complex structures (including links and interconnected structures). The results of this study showed that combination of different tools allows us to gain appropriate information through a comprehensive and fast way[8].

During Vultur studies, evaluation of five websites of Romanian economics colleges was conducted using WIA method. In this study, WIA has five main categories. Findings from this research showed that Romanian economics colleges websites were in average condition from the perspective of accessibility, in a very desirable condition in terms of velocity, in a desirable condition in terms of navigation, in the average condition from the content viewpoint, and in the average condition in terms of confidence[9].

Mustafa et al. and Islam et al. also used two automated online tools i.e. HTML toolbox and web page analysis to evaluate the websites of the universities in Jordan[10, 11]. The result of the evaluation was used to provide suggestions for the improvement of the usability of websites[10].

In a study conducted by Eyadat and Lew[12], Fernandez et al.[13] and Hashemian[14], the automated tool of Test Accesibilidad Web (TAW) from Information and Communication Technology Development Center was used[15], which is an online service that analyzes websites for adaptability to WGAG 2.0.

Arif et al. collected the data they needed for their study using Google search engine and Majestic SEO in July 2012 wherein the number of pages was indexed from URL and the keyword was considered as a parameter. The findings suggested that Malaysian state universities did not have a significant impact on the virtual world[16].

In the study of Ismailova and Kimsanova, "accessibility and usability" of university websites in Kyrgyzstan Republic were reviewed based on the accepted standards and usability using online evaluation tools. According to the results, the usability rank was low on most university websites[17].

Kaur et al. concentrated on ways to find all the possible parameters in designing a website referring to some of the key universities in Punjab. Using the four automated tools of SEO the websites under the study were evaluated and the comparative results of various parameters were stated. The findings of this study indicate that usability of automated tools for website performance can help save time and improve the quality of website designing using various parameters, performance, requests, load time, page size, user experience, mobile, SEO, and security[18].

A review of literature suggests that website evaluation can be done using specific methods. This evaluation can be based on expert use. An automated tool-based evaluation uses an automated tool to determine the internal (or fundamental) features of the website. Features like number of HTML pages, page size, number of broken or bad links and other technical defects on web pages are discovered through these tools[19, 20].

## 3. Research Methodology

This is an applied research with a webometric method with a descriptive cross-sectional approach, which was conducted in January 2016. In the following, all stages of the implementation will be described separately.

**3.1. University Selection:** The present study consisted of 75 university websites that in this survey, firstly, name of the 19 domains along with the world ranks of the university websites that ranked below 1000 in the webometric ranking results were extracted randomly from the Webometric website[C]. Subsequently, 56 universities that ranked poorly in Webometric ranking and ranked between 1,000 and 22,500 were randomly extracted.

**3.2. Extracting effective metrics in SEO:** Among the related works in the field of SEO and the most important parameters and metrics of SEO that should be considered in designing website (On-Page SEO) and after designing website (Off-Page SEO) that have a great impact on SEO, 38 top metrics were extracted and were considered as an indicator. In Table 1, names of SEO metrics examined in this study are listed.

**Table 1. SEO Metrics for the research**

| SEO Metrics for the research | |
|---|---|
| Alexa Rank | Page size |
| The number of back links | Encoding |
| Total links | Robot.txt File |
| Number of internal link | Sitemap |
| Number of External link | Responsive |
| Number of broken links | Social Media |
| Trust flow (Link Quality) | The number of pages indexed in Google |
| Number of requests | language(English) |
| Load time | DocType |
| Tag H1 | Page 404 |
| Number of IMG tags without attribute alt | Gzip |
| Tag iframe | Referring domains |
| Embed- Object Tag | Referring IPs |
| Number of HTML error | Security |

[C] www.webometrics.info

| | |
|---|---|
| Number of HTML warning | Performance |
| Number of CSS error | Accessibility |
| Number of CSS warning | Page Rank (PR) |
| number of character use in Title Tag | Domain Authority (DA) |
| number of characters use in Meta Description | Page Authority (PA) |

### 3.3. Extracting the scores of the selected universities in each parameter of SEO

To extract the scores of the universities under the study in each SEO parameter, 16 automated SEO tools were used whose names are listed in Table 2.

**Table 2. The Automated SEO Tools for the Research**

| **The Automated SEO Tools for the Research: SEO Metrics for the research** | |
|---|---|
| **majestic.com:** Referring IPs, Referring domains, Backlink, Trust flow | **websiteseochecker.com:** Domain Authority, Page Authority |
| **gtmetrix.com:** Number of requests, Load time, Page size | **webconfs.com:** Alexa Rank |
| **website.grader.com**: Responsive, Security, Performance | **sitecheckup.ir:** Encoding |
| **validator.w3.org:** HTML error, HTML warning | **seocentro.com:** Social Media |
| **jigsaw.w3.org/css-validator:** CSS error, CSS warning | **prchecker.info:** Page Rank |
| **qualidator.com:** Accessibility | **tools.hostiran.net:** Embed- Object Tag, Page 404 |
| **smallseotools.com:** Total links, internal link, External link | **seorch.eu:** iframe Tag |
| **seositecheckup.com:** broken links,H1 Tag, IMG | **Google search engine:** The number of pages indexed |
| tag without alt, number of character use in Title Tag, number of characters use in Meta Description, Robot.txt File, Sitemap, DocType, Gzip | in the Google search engine |

For example, using Majestic tool, metrics of the number of referring domains, the number of referring IP, the number of Back links, and link quality [Trust flow]; and using the Google search engine and Site: URL command, the number of web pages indexed in Google search engine for any website were extracted.

It should be noted that since webometric  was the main reference of quality metrics of university website in this study, if a specific tool was mentioned in this site to extract quality metrics; this study used the same tools.

## 3.4. Data Analysis and Mining

**3.4.1. Two-Variable Dispersion Distribution Chart Analysis:** In this analysis, first the data extracted for 38 SEO parameters examined in the study population were separated in terms of normal, abnormal, and those which can be normalized by approximation. Then conversion operation was performed for the parameters that their data could be normalized by approximation using logarithm, because one of the ways of normalization is to calculate the logarithm of the data. To do this, the data extracted for each SEO parameter was examined and if there were any zero (0) values, initially, the total data of the desired SEO parameter was summed with 0, 1 and then their logarithm was calculated. In order to evaluate the impact and performance of each SEO parameter on the university websites rank, the method of drawing two-variable dispersion distribution chart was used. It should be noted that for SEO parameters with normal data, two-variable dispersion distribution chart was drawn normally without logarithm calculation but it was not drawn for other SEO parameters with abnormal data. To perform this step of the research, SPSS Statistics 24 was used.

**3.4.2. Association Rules Analysis:** Using SPSS Modeler 18, using dimension reduction method, and feature selection, 14 out of 38 SEO parameters under the study with the highest impact were identified and extracted then among the 14 parameters of SEO extracted, 4 parameters of security, language, tags Title and Iframe whose data values were zero and one[D] and had less impacts compared to the 10 other parameters were manually excluded from this analysis and a total of 10 more important SEO parameters such as Page Rank, Alexa Rank, Back links, External links, Quality links, Domain Authority, Page Authority, the number of indexed pages, the number of referring domains, and the number of referring IP were studied. Finally, using model association and association rule method, the analysis of the 10 desired parameters for the website of the universities under the study was performed.

[D] True & False

### 3.5. Finding High-Impact Parameters and Offering Solutions

After the performed investigations, the most important parameters in SEO and their impact on webometric rankings were known and suggestions were made to improve the status of university websites in terms of content and design.

## 4. Findings and Discussion

In general, our aim in this study was to answer two main questions that are answered to each in details in the following.

4.1. How much do the SEO parameters of a university website affect university website rank ?

To answer this question, analysis of the two-variable dispersion distribution chart was used. By examining the data, the impact of each SEO parameter on the Webometric rank was obtained and the results were mentioned in Table 3.

**Table 3. The impact of each SEO parameter on Webometric rank with two-variable dispersion distribution chart**

| SEO Parameter | Logarithm | Impact of the Parameter on Webometric Ranking |
|---|---|---|
| Number of pages indexed | ✓ | 0.662 |
| Alexa Rank | ✓ | 0.576 |
| Backlink | ✓ | 0.574 |
| Referring domains | ✓ | 0.490 |
| Referring IPs | ✓ | 0.462 |
| Domain Authority | - | 0.455 |
| Page Rank | ✓ | 0.390 |
| Page Authority | - | 0.335 |
| Social Media | ✓ | 0.201 |
| External link | ✓ | 0.162 |
| Trust flow | ✓ | 0.146 |
| Accessibility | - | 0.050 |
| CSS error | ✓ | 0.034 |
| Performance | - | 0.014 |
| IMG tag without alt | ✓ | 0.012 |
| HTML warning | ✓ | 0.012 |
| number of character use in Title Tag | ✓ | 0.012 |
| HTML error | ✓ | 0.010 |
| number of characters use in Meta Description | ✓ | 0.007 |
| Page size | ✓ | 0.005 |
| Load time | ✓ | 0.004 |
| CSS warning | ✓ | 0.004 |
| Total links | ✓ | 0.003 |
| Number of requests | ✓ | 0.002 |
| internal link | ✓ | 0.001 |
| broken links, H1 Tag, IFrame Tag, Embed-Object Tag, Encoding, Robot.txt File, Sitemap, Responsive, Language, DocType, Page 404, Gzip, Security | - | - |

In the table above, the parameters in the logarithm column that have a tick mark (✓) in the front (Alexa rank, back links, Page Rank, etc.) means that the values of these parameters were not normal and the data have been normalized by approximation using the conversion operation and logarithm. Finally, a two-variable dispersion distribution chart can be created as shown in Figure 1 for the logarithm of the desired SEO parameter and for the rank of the website of the universities.

According to data normality for the parameters of performance, availability, Page Authority, and Domain Authority, logarithm was not calculated and their two-variable dispersion distribution chart could be normally created for the desired SEO parameter and for the website rank of the universities.

In addition, since the values of the data for the parameters of broken links, H1 tags, robot.txt File, Site map, responsive, language, Doctype, Page 404, Iframe tags, Embed-object tag, Encryption, Gzip compression and security were abnormal; they could not be normalized by approximation.

For example, based on the findings, Figure 1 is an example of the created charts that shows the relationship of back link logarithm with world Webometric rank. As clearly shown, the more is the number of back links, the lower and the better is the website rank. Based on the findings, the impact of back link logarithm with Webometric rank is 0.574 and highlights the immense impact of back link on Webometric rank.

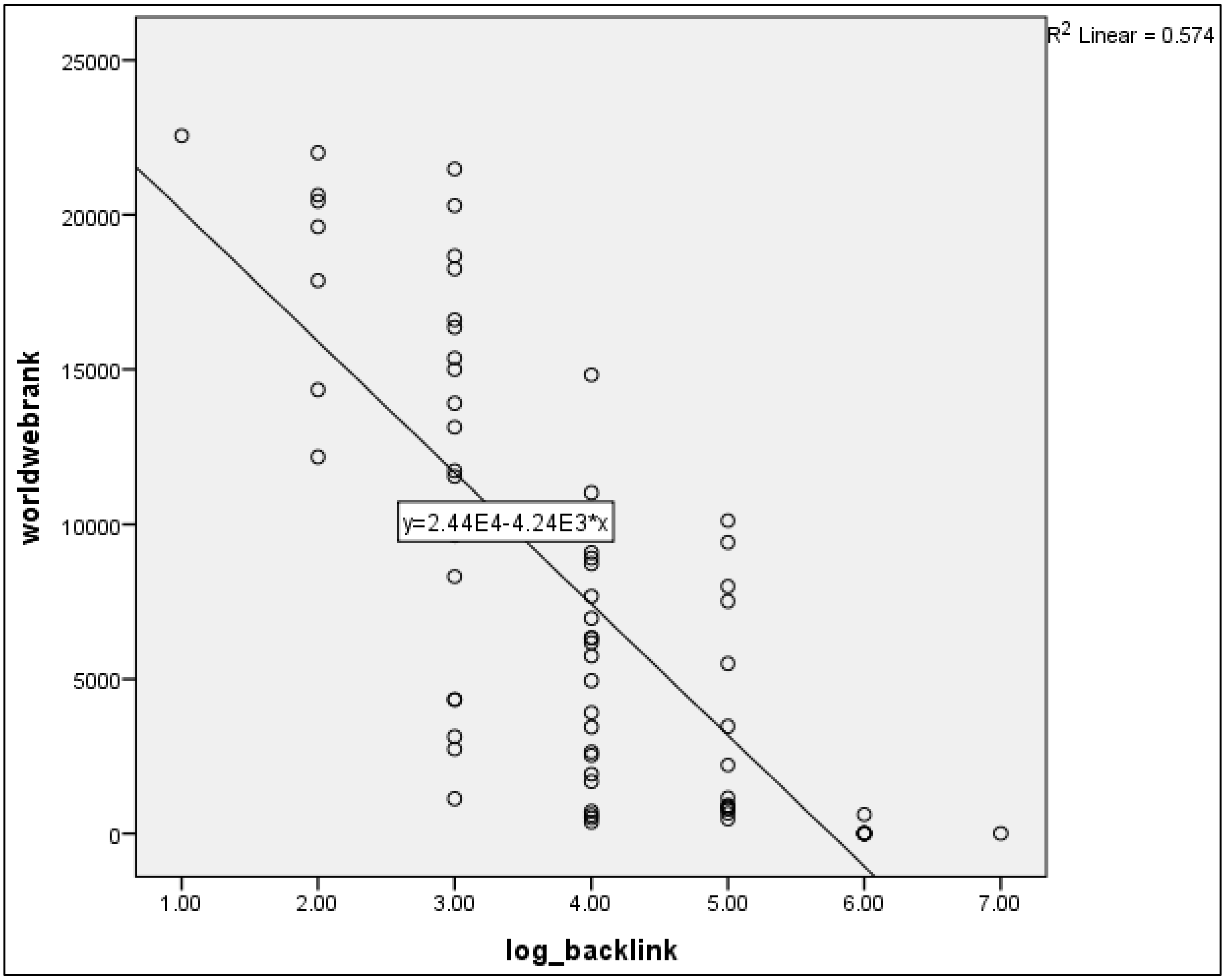

**Figure 1. Relationship of Log-Back Link with World Webometric Rank**

According to the findings of the first question, the metrics of Alexa rank, back link, PageRank, external links, link quality, the number of indexed pages, social network, the number of referring domain and IP, and authority of page and domain significantly affect the visibility of the website. Based on these findings, metrics for the internal optimization of search engines should be considered by university website designers because all these metrics together will have a significant impact on the visibility of the website in search engines.

It should be noted that parameters affecting the external optimization of the website also complement its internal optimization and increase the visibility and the rank of the web page. In the study of Jalal et al., most of these parameters were studied and similar findings were found in this field[21]. Therefore, complying with the metrics of On-Page SEO and Off-Page SEO, metrics of Alexa rank, PageRank, Page Authority, Domain Authority, and website performance will also improve.

4.2. What important rules are extracted regarding SEO parameters and university websites rank using association rules?

To answer this question, first, the feature selection analysis was used by which 10 out of 38 parameters with the highest impact were identified and extracted. After performing feature selection method, association rule analysis was used. According to the findings, 30 interesting rules were extracted by confidence and 30 interesting rules were extracted by rule supports that some of the rules were the same between these two items (confidence and rule supports). In Table 4, the most important extracted rules, which had the highest coefficient of confidence, have been listed.

**Table 4. The Most Interesting Extracted Rules**

| Status | Forecast | Sorted By Confidence (%) | Rule Supports (%) |
|---|---|---|---|
| PR > 7 | Webometric ranks ≤ 4,511 | 100 | 12 |
| DA> 75 | Webometric ranks ≤ 4,511 | 100 | 12 |
| PA> 76 | Webometric ranks ≤ 4,511 | 100 | 10.67 |
| Trust flow > 71 | Webometric ranks ≤ 4,511 | 100 | 5.33 |
| 53 ≤ Trust flow < 71 | Webometric ranks ≤ 4,511 | 90 | 12 |
| 5 ≤ PR < 7<br>36 ≤ Trust flow < 53 | Webometric ranks ≤ 4,511 | 100 | 8 |
| PR > 7<br>118 ≤ Number of external links < 177 | Webometric ranks ≤ 4,511 | 100 | 5.33 |
| PR ≤ 1<br>Number of external links ≤ 59<br>Trust flow ≤ 18<br>Indexed pages ≤ 4,800,005 | Webometric ranks > 18,041 | 61.54 | 10.67 |

For example, based on rules extracted in row 1 in Table 4, if the university website has a PageRank greater than 7, website rank of that university is less than or equal to 4,511. Based on confidence, 100 percent of the university websites that have a PageRank greater than 7 will have a Webometric rank less than 4,511 and based on rules support, 12 percent of the universities surveyed have a PageRank greater than 7 and the Webometric rank is less than or equal to 4,511.

According to the findings of association rules analysis, among 10 parameters considered in this analysis, for the 8 parameters of PageRank, Alexa rank, Domain Authority, Page Authority, links quality, the number of referring IP, the number of indexed pages, and external links, rules were extracted that shows these parameters have a major influence on rules. According to the findings of rules support in the analysis of association rules, a maximum of 12 percent of the universities surveyed in the desired SEO parameters have high ranks. Thus, out of the 75 universities surveyed in the study, only top universities of the world (including Harvard, Stanford, Massachusetts, Berkeley, etc) could achieve a much higher rank than other universities in the world in terms of compliance with the eight SEO parameters examined in this analysis.

## 5 . Conclusion & Future Works

SEO is essentially aimed at improving the visibility of a website in search engine results page. To achieve this goal, a website must be optimized in terms of several internal and external parameters. As visibility (V) and presence (S) are among the four indicators used to calculate the Webometric features, they can be used in implementing SEO. To improve the rank in Webometric, the university should improve On-Page SEO parameters on the website of the university by implementing SEO techniques. On-Page SEO is related to what you write and access on your website such as providing proper content, proper keywords, title tag, meta tag, etc.

After improving On-Page parameters of SEO on the website, Off-Page SEO parameters must be improved as well. Off-Page SEO is also related to link building, getting a site into large directories and search engines. Following the improvement of Off-Page SEO parameters, indicator of presence (S), followed by the indicator of visibility (V) of university website will increase and eventually university rank in Webometric can be increased.

An important point in the university websites ranking is that according to a Webometric site report on January 2016, around 70% of the university websites ranking criteria are directly related to web criteria. Accordingly, considering that the visibility indicator (V) with 50% has the highest coefficient, size indicator (S) has a coefficient of 20%, thus a combination of these two indicators can increase the total score of Webometric up to 70%. These two indicators directly and indirectly are related to SEO and according to the results of this study, it can be said that SEO can affect up to 70% of the university websites ranking.

Studies of previous Webometric reports indicate that the coefficient of the SEO indicators impact has gradually diminished and research indicators have increased. For future works, not only SEO criteria can be extended but also the impact of other web standards on web site rankings can be examined. Testing the other data mining methods or their combination may lead to newer and more accurate findings.